\documentclass[pra,aps,superscriptaddress, twocolumn,showpacs,floatfix]{revtex4}
\usepackage{amsfonts}
\usepackage{amsmath}
\usepackage{bm}
\usepackage{epsfig}
\usepackage{mathrsfs}

\usepackage{array,dcolumn}

\begin{document}

\title{Low-lying dipole response in the stable  
$^{40,48}$Ca nuclei with the second random-phase approximation}

\author{D. Gambacurta}
\address{Dipartimento di Fisica e Astronomia and INFN, Via Santa
  Sofia 64, I-95123 Catania, Italy}
\address{Grand Acc\'el\'erateur National d'Ions Lourds (GANIL), CEA/DSM-CNRS/IN2P3, Bvd Henri Becquerel, 14076 Caen, France}

\author{M. Grasso}

\address{Institut de Physique Nucl\'eaire,
 Universit\'e Paris-Sud, IN2P3-CNRS, F-91406 Orsay Cedex, France}

\author{F. Catara}
\address{Dipartimento di Fisica e Astronomia and INFN, Via Santa
  Sofia 64, I-95123 Catania, Italy}

\begin{abstract}
Low-energy dipole excitations are analyzed for the stable isotopes $^{40}$Ca and
$^{48}$Ca in the framework of the Skyrme-second random-phase approximation. The
corresponding random-phase approximation calculations provide a negligible
strength distribution for both nuclei in the energy region from 5 to 10 MeV. The
inclusion and the coupling of 2 particle-2 hole configurations in the second
random-phase approximation lead to an appreciable dipole response at low
energies for the neutron-rich nucleus $^{48}$Ca. The presence of a neutron skin
in  the nucleus $^{48}$Ca would suggest the interpretation of the low-lying
response in terms of a pygmy excitation. The composition of the excitation modes
(content of 1 particle-1 hole and 2 particle-2 hole configurations), their
transition densities and their collectivity (number and coherence of the
different contributions) are analyzed. This analysis indicates that, in general, these
excitations cannot be clearly interpreted in terms of oscillations of the
neutron skin against the core with the exception of the peak with the largest
$B(E1)$ value, which is located at 9.09 MeV. For this peak the neutron transition density
dominates and the neutron and proton transition densities oscillate out of phase
in the internal part of the nucleus leading to a strong mixing of isoscalar and
isovector components. Therefore, this state shows some features usually
associated to pygmy resonances. 
\end{abstract}

\vskip 0.5cm \pacs{21.60.Jz,21.10.Re} \maketitle

\section{Introduction}

The evolution 
of the low-lying dipole response as a function of the isospin asymmetry has
been 
extensively analyzed both experimentally and theoretically in several stable and
unstable nuclei.
A recent review about 
the main theoretical results and experimental measurements can be found in Ref.
\cite{paar}. 
Measurements on the halo nuclei $^6$He, $^8$He, $^{11}$Li, $^{12}$Be, $^{19}$C
and  
$^{8}$B (Refs. \cite{uno,paar} and references therein) have shown the presence
of a significant dipole strength at very low energy (for instance, 
in the case of $^{11}$Li, below 4 MeV). In these light nuclei, the 
development of a strong dipole response at low energies can be 
related to the extremely small value of the separation energies (the systems are
very weakly bound and the last occupied neutron states are close to the
continuum states). These low-energy states are not collective excitations 
and have mainly a single-particle character \cite{cdv96}. The same kind of
picture
(individual excitations of the weakly bound last 
occupied neutron states) is currently provided by most of the available
theoretical models 
in the description of the low-lying excitations in neutron-rich Oxygen 
isotopes. Low-energy dipole excitations have been observed for the isotopes
$^{16-22}$O and the corresponding 
data are reported in Refs. \cite{due}. From the experimental point of view, the
character of these excitations (collective or 
single-particle) has not yet been clearly elucidated. 

For heavier stable and unstable nuclei, the development of a pygmy dipole
response in neutron-rich systems 
is currently related to the formation of a thick neutron skin at the surface of
the nucleus: the 
low-lying dipole modes are interpreted in terms of oscillations of the skin
against the core 
composed by both neutrons and protons. Experimentally, low-lying $E_1$ states
have been measured in 
several medium-mass and heavy nuclei such as, for example, 
$^{40,44,48}$Ca \cite{tre,tre-1}, the tin isotopes $^{112}$Sn 
\cite{sei}, $^{116,124}$Sn \cite{sette}, $^{130}$Sn and $^{132}$Sn
\cite{dieci}, 
$^{204,206,207,208}$Pb \cite{otto}, and $^{68}$Ni \cite{milano}. 

From the theoretical point of view, pygmy resonances have been analyzed with
several models. Some examples 
are 
the relativistic and non-relativistic (Q)RPA approach (see, for instance, Ref.
\cite{paar} and 
references therein, and Refs. \cite{peru}-\cite{Martini}), the particle-phonon-coupling
models (Ref. \cite{paar} and references therein), a semiclassical
coupled-channels approach for Sn isotopes \cite{lanza,vitturi}, hydrodynamical
models \cite{mohan}, the phonon-damping model \cite{dang}, and the so-called
Extended Theory of Finite Fermi 
Systems (ETFFS) \cite{tertychny}.  
Discrepancies among the different theoretical predictions are  
found concerning in particular 
the collective character and the fragmentation of the low-lying modes. 

The widths and the fragmentation of the excited modes 
in a many-body system cannot be described within the 
standard random-phase approximation (RPA) which can only account for the
so-called Landau damping (related to single-particle degrees of freedom). It is
well known that,  
to describe widths and fragmentation, the single-particle degrees of freedom
have to be coupled with more complex configurations (collective coordinates or
multiparticle-multihole configurations)
within a beyond mean-field model.  
Among the different beyond mean-field models that allow one to describe, at least
partially, the width and the fragmentation of the excitation modes, the second
random-phase approximation (SRPA) is a powerful theoretical tool where the
coupling with 2 particle- 2 hole ($2p2h$) configurations is included within an
RPA-like formalism. In this way, the so-called spreading widths can be described
together with the Landau damping. Escape widths are missing if the coupling to
the continuum is not included. To include higher multiparticle-multihole configurations different 
directions may be followed \cite{pillet,loiudice,grasso}.
 
Due to the heavy numerical effort required,  
the SRPA equations have been often solved resorting to some approximations, 
namely the SRPA equations have been reduced to a simpler 
second Tamm-Dancoff model (i.e. the matrix $B$ is put equal to zero, see for
instance \cite{ho,kn,ni1,ni2}) 
and/or the equations have been solved with uncorrelated 
$2p2h$ states in the so-called diagonal 
approximation \cite{ad,sc1,sc2,dr1,dr2,yann2,yann3}.
Recently, full SRPA calculations have been performed for some O and Ca isotopes
\cite{roth, gamba2010-1}. In particular, in Ref. \cite{gamba2010-1} calculations
with the density-dependent Skyrme interaction have been performed adopting two
currently used approximations for treating the
rearrangement terms of the 
 residual interaction 
appearing in beyond-RPA matrix elements. The two approximations consist in
either neglecting these rearrangement terms or treating them with the standard
RPA procedure. Important differences have been found 
between the corresponding two sets of results. The same authors have addressed
this point in 
a more recent work \cite{gamba2010-2} where a procedure to derive the 
expressions of all the rearrangement terms within the SRPA framework has been
presented and applied to calculations for the nucleus $^{16}$O. In this first
application, the importance of the proper treatment of the 
rearrangement terms in SRPA for the description of the fragmentation of the
excited modes has been shown. 

In this work, we employ the implemented code where the full rearrangement terms
have been included to treat the low-lying excitation spectrum of the stable
isotopes $^{40,48}$Ca within the Skyrme-SRPA model. 
Medium-mass Ca isotopes are chosen as intermediate cases between light nuclei
where the low-lying dipole excitations have mainly a single-particle character
and heavier nuclei like Sn and Pb isotopes. Ca isotopes are expected to be
interesting cases where 
the low-energy modes could eventually start to be more collective (with respect to light nuclei) and 
the nature of the low-lying excitations in terms of collectivity and 
fragmentation may be investigated. 

The paper is organized as follows. In Sec. II the main formal aspects of the
SRPA model are briefly recalled. In Sec. III the low-lying strength
distributions are analyzed for the nuclei $^{40,48}$Ca 
and the transition densities associated to some states are displayed.
In Sec. IV some comments about the spurious state are presented. 
We draw our conclusions in Sev. V.

\section{Brief summary of the formal aspects of SRPA}

The SRPA equations are known since many years and have been derived by following
different procedures, such as the equations-of-motion method \cite{yannou}, the
small-amplitude limit of the time-dependent density matrix method
\cite{tohyama,lacroix}, and a variational procedure introduced by da Providencia
\cite{pro}. The main properties are also recalled in more recent works
\cite{roth,gamba2010-1}. 

The excited states in SRPA are superpositions of 1 particle-1 hole ($1p1h$) and
$2p2h$ configurations. The SRPA equations can be written in the compact form

\begin{equation}\label{eq_srpa}
\left(\begin{array}{cc}
  \mathcal{A} & \mathcal{B} \\
  -\mathcal{B}^{*} & -\mathcal{A}^{*} \\
\end{array}\right)
\left(%
\begin{array}{c}
  \mathcal{X}^{\nu} \\
  \mathcal{Y}^{\nu} \\
\end{array}%
\right)=\omega_{\nu}
\left(%
\begin{array}{c}
  \mathcal{X}^{\nu} \\
  \mathcal{Y}^{\nu} \\
\end{array}%
\right),
\end{equation}
where:
\begin{displaymath}
\mathcal{A}=\left(\begin{array}{cc}
  A_{11} & A_{12} \\
  A_{21} & A_{22} \\
\end{array}\right),
\mathcal{B}=\left(\begin{array}{cc}
  B_{11} & B_{12} \\
  B_{21} & B_{22} \\
\end{array}\right),
\end{displaymath}
\begin{displaymath}
\mathcal{X}^{\nu}=\left(\begin{array}{cc}
  X_{1}^{\nu} \\
   X_{2}^{\nu} \\
\end{array}\right),
~~~~\mathcal{Y}^{\nu}=\left(\begin{array}{cc}
  Y_{1}^{\nu} \\
   Y_{2}^{\nu} \\
\end{array}\right).
\end{displaymath}

In the above equations, '1' and '2' stand for $1p1h$ and $2p2h$, respectively.
Thus,  
$A_{11}$ and $B_{11}$ represent the usual RPA matrices, whereas the matrices 
$A_{12}$ and $B_{12}$ couple $1p1h$ with $2p2h$ configurations and the matrices 
$A_{22}$ and $B_{22}$ couple among themselves $2p2h$ configurations. The
detailed expressions of these matrices can be found for example in Ref.
\cite{gamba2010-1}. If a density-dependent interaction like the Skyrme 
force is employed, rearrangement terms appear in the residual interaction. 
 The usual RPA rearrangement terms appear 
in the matrices $A_{11}$ and $B_{11}$. New types of 
rearrangement terms have been obtained for the other matrix elements in Ref.
\cite{gamba2010-2} within a variational derivation of the SRPA equations. The
expressions of these rearrangement terms are reported in Ref. \cite{gamba2010-2}
and are used 
in this work. 

Before analyzing the results, 
we recall here two main properties of SRPA: the quasiboson approximation (QBA)
is adopted in standard SRPA and is adopted here; the energy-weighted sum
rules (EWSRs) are satisfied in SRPA as demonstrated formally in Ref.
\cite{yannou} and verified numerically 
 in Ref. \cite{gamba2010-1}.

\section{Results for $^{40,48}$Ca}

Both Ca isotopes are stable but a neutron skin has been measured experimentally
in $^{48}$Ca by proton and electron scattering experiments \cite{caskin}. 
The proton (neutron) radii found within the SGII-Hartree-Fock model are equal to 
3.37 (3.32) fm and 3.41 (3.55) fm for the nuclei $^{40}$Ca and $^{48}$Ca, 
respectively. 
The
experimental low-lying dipole response has been recently analyzed in the two
isotopes \cite{tre} and the development of a low-energy strength, between 5 and
10 MeV, has been observed in 
$^{48}$Ca.
 \begin{figure}
\begin{flushleft}
 
\epsfig{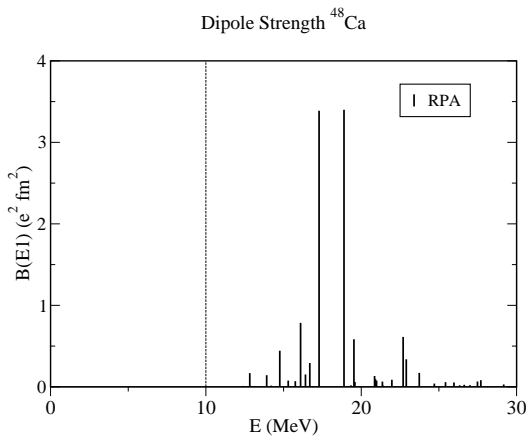}
\caption{\small RPA dipole strength distribution for $^{48}$Ca. }
\label{RPA-Ca48}
\end{flushleft}
\end{figure}
From the theoretical point of view, it has been found that relativistic and
non-relativistic (Q)RPA models are not able to well describe the low-lying
response in 
$^{48}$Ca because they either do not provide the good excitation energies (too
high energies) or do not  predict the experimental fragmentation of the peaks.
For example, in recent calculations performed with the relativistic RPA model no
strength has been found in the response below the excitation energy of 10 MeV
\cite{vret}. 
The same kind of results is obtained in Skyrme SGII-RPA calculations (Fig.
\ref{RPA-Ca48}).  
On the other hand, 
a reasonable agreement (energies and fragmentation) with the experimental
results has been found within the ETFFS model \cite{tre-1,tertychny} where a quasiparticle-phonon 
coupling is included.

We perform SRPA calculations in spherical symmetry for the two isotopes
$^{40}$Ca and $^{48}$Ca with the Skyrme interaction SGII. The technical details
of these calculations are reported in Ref. 
\cite{gamba2010-1}. Differently from Ref. \cite{gamba2010-1},  
the full rearrangement terms \cite{gamba2010-2} are used
here in the SRPA matrices.
Because of the zero-range of the Skyrme interaction, a natural energy cutoff is not provided. Different  procedures to treat this problem  may be envisaged for future
studies (see, for instance, the exploratory work presented in Ref. \cite{moghrabi}). In this work, we 
have introduced an energy cutoff ($ECUT$) on the $2p2h$ configurations. By varying it from 40 to 60 MeV 
 we have
verified that a
reasonable stability of the results
is achieved around a cutoff of 50-55 MeV. 
The total $B(E1)$  and EWSRs values, integrated up to an energy of 10 MeV, are
shown in Table \ref{tab-0} for the isotopes 
$^{40}$Ca and $^{48}$Ca as a function of the energy cutoff $ECUT$ on the $2p2h$ configurations.

\begin{figure}
\begin{flushleft}
\epsfig{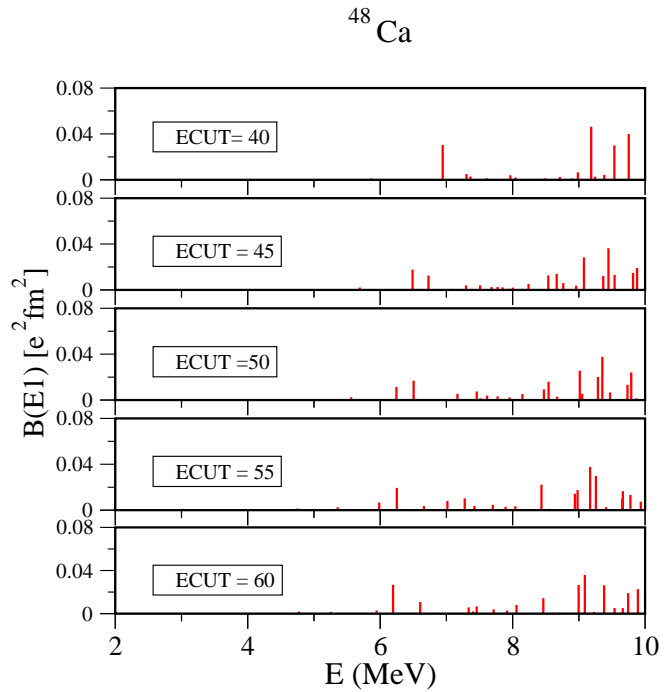}
\caption{\small (Color online) SRPA dipole strength distribution for the $^{48}$Ca
isotope
for increasing values of the energy cutoff $ECUT$ (MeV)
 on the $2p2h$ configurations included in the calculations. }
\label{Ca48}
\end{flushleft}
\end{figure}

\begin{figure}
\begin{flushleft}
\epsfig{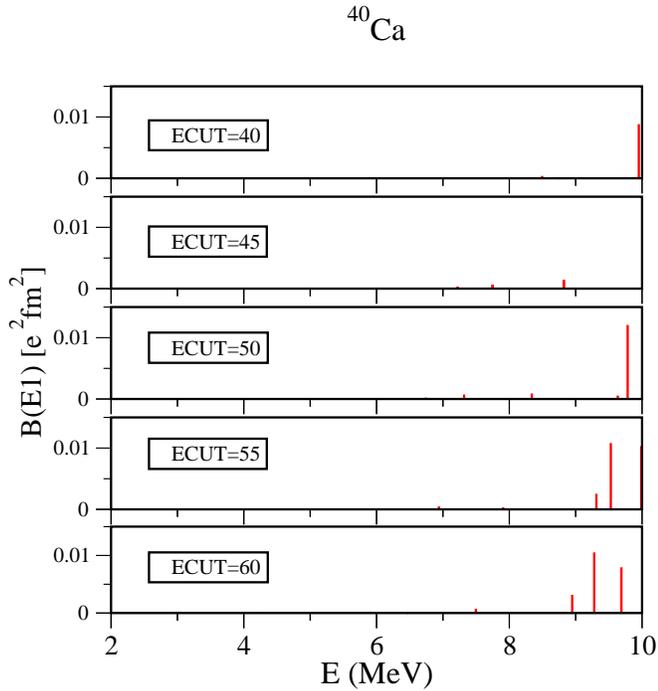}
\caption{\small (Color online) As in Fig. \ref{Ca48} but for the $^{40}$Ca
isotope. Please note that a different scale has been used in the ordinate with
respect to Fig. \ref{Ca48}.}
\label{Ca40}
\end{flushleft}
\end{figure}

The $B(E1)$ distributions for
different choices of the energy cutoff are plotted in Fig. \ref{Ca48} up to an excitation
energy
of 10 MeV for the nucleus $^{48}$Ca.
The employed transition operator is

\begin{equation}\label{trans-operator}
 F_{10}=e_p\sum_{i=1}^{Z} r_i Y_{10}(\Omega_i)-
e_n\sum_{i=1}^{N} r_i Y_{10}(\Omega_i)
\end{equation}
where $e_p$ and $e_n$ are the kinematic charges, $e_p=Ne/A$ and  $e_n=Ze/A$,
respectively.
One can observe that the results do not
change strongly starting from a cutoff of 45 MeV. In what follows, we will analyze the
results obtained 
with a cutoff of 60 MeV (bottom panel of Fig. \ref{Ca48}). 
\begin {table} 
  \begin{center}
\begin{tabular}{|c|cc|ccc|}
\hline
&$^{48}$Ca&&&$^{40}$Ca&\\
\hline
$ECUT$ &	$\sum B(E1)$&	EWSRs&&  	$\sum B(E1)$&	EWSRs \\
\hline             
40&	0.184&	1.623&&  	0.009 &   0.091\\
\hline             
45&	0.218&	1.895&& 	0.002 &   0.022\\
\hline              
50&	0.226&	1.944&&  	0.015 &   0.139\\
\hline             
55&	0.240&	2.049&&  	0.025 &   0.237\\
\hline              
60&	0.230	&1.964 && 	0.023 &   0.211\\
\hline 
\end{tabular}\caption {\label{tab-0} Total $B(E1)$ ($e^2 fm^2$) and EWSRs ($e^2
fm^2 MeV$) integrated up to 10 MeV as a function of the energy cutoff $ECUT$
($MeV$) on the $2p2h$ configurations for $^{48}$Ca and $^{40}$Ca.}

\end{center}
\end{table}
These results are qualitatively of the same type as those found in Ref.
\cite{tre-1} with the ETFFS approach which is also a beyond-mean-field model where
the coupling is done with collective phonons instead of $2p2h$ configurations
(as is done in SRPA). We can compare the location of the theoretical peaks with the experimental distribution (see, 
for instance, Fig. 2 of Ref. \cite{tre-1}). We can distinguish two regions: from 6 to 8 MeV and from 
8 to 10 MeV. Experimentally, the highest peak in the first region is found at $\sim$ 7 MeV whereas in our case we have 
several small peaks between 6 and 8 MeV and the highest peak is located around 6.2 MeV. In the interval 
between 8 and 10 MeV, our response is more fragmented than the experimental one. Experimentally, peaks are found around 
8.5, 9 and 9.5 MeV. Our highest peak is located at $\sim$ 9.1 MeV. 
Finally, 
we have evaluated the response in the nucleus $^{40}$Ca. Experimentally, a
negligible strength has been found for this nucleus between 5 and 10 MeV 
(Fig. 2 of Ref. \cite{tre-1}). Our
results for $^{40}$Ca are displayed in Fig. \ref{Ca40} for cutoff values varying from 40 up
to 60 MeV. Some peaks are actually found at low energy but the corresponding strength is much
lower than in $^{48}$Ca (notice the different scales in the two figures).

\begin{figure}
\begin{flushleft}
\epsfig{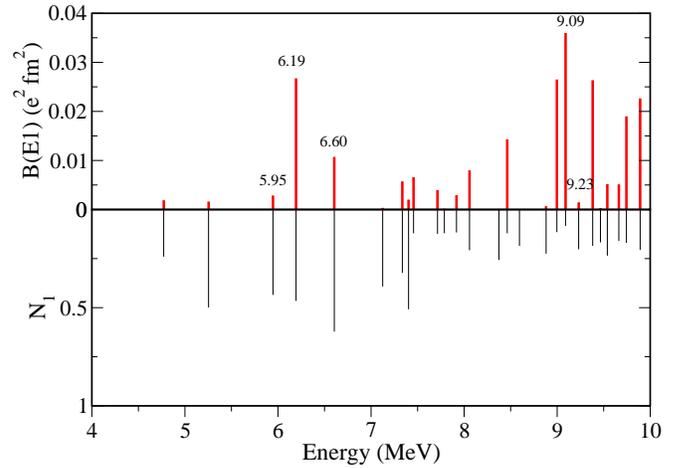} 
\caption{\small (Color online) For each state the $B(E1)$ value corresponding
to a cutoff of 60 MeV (upper panel) and the total $1p1h$ contribution $N_1$ 
 to the norm of the state defined in Eq. (\ref{norm}) (lower panel),  are shown for $^{48}$Ca. }
\label{Ca48N1}
\end{flushleft}
\end{figure}

An interesting information that can be analyzed in connection with the strength
distribution is the composition of the excitation modes in terms of $1p1h$
 and $2p2h$ configurations. By extracting the expression of $N_1$ from the SRPA normalization condition, 
\begin{displaymath}
 \sum_{ph}(\mid X_{ph}^{\nu}\mid^2-\mid Y_{ph}^{\nu}\mid^2)+
\sum_{p<p',h<h'}(\mid X_{php'h'}^{\nu}\mid^2-\mid Y_{php'h'}^{\nu}\mid^2)
\end{displaymath}
\begin{equation}\label{norm}
 = N_1+N_2=1,
\end{equation}     
we plot in Fig. \ref{Ca48N1} the $B(E1)$ values corresponding
to a cutoff of 60 MeV (upper panel, same as in bottom panel of Fig. \ref{Ca48}) and the
quantity $N_1$  (lower panel) for each excitation of 
the nucleus $^{48}$Ca. One observes that all the excitations present a mixing of
$1p1h$ and $2p2h$ configurations. Those which have the highest $1p1h$ content
(around 50\%) may be interpreted as excitations that already exist in the RPA spectrum at higher
energies (the first excitations in RPA are located around 11 MeV) and that are
shifted down to lower energies due to the
coupling with $2p2h$ configurations. In the SRPA spectrum we also see several states
which present a dominant 
$2p2h$ nature and show a relatively large $B(E1)$ 
despite their very low content of $1p1h$ configurations (see for instance
the energy region around 9 MeV). 

As already mentioned in the Sec. I, low-energy excitations in light nuclei
are mostly single-particle excitations and cannot be interpreted as a collective
motion of a skin against a core. In Ca isotopes, which are intermediate cases
between light and heavy nuclei, the nature of these excitations has not yet been 
clearly elucidated. In order to have a deeper insight into the properties of these low-energy modes we consider in
more detail 
the states located at 5.95, 6.19 and 6.60 MeV as well as those located at 9.09 and 9.23 MeV.
We can see that the first three states have a quite large and similar
$1p1h$ component but show a very different $B(E1)$ value. In particular, one notices that the
state located 
at 5.95 MeV has almost no strength. By comparing among themselves the states with energies 6.19 and 6.60 MeV, 
one observes that $N_1$ 
is larger while the $B(E1)$ value is about one half 
in the second state with respect to the first one. A different description is provided for the states lying at 9.09 and 9.23 MeV which are both 
almost entirely composed by $2p2h$ configurations. 
In spite of the fact that the $1p1h$ content is higher in the second state with 
respect to the first one, we notice that the second state has almost no strength while  
the first state is the most collective in the low-lying spectrum. 

\begin{figure}
\begin{flushleft}
\epsfig{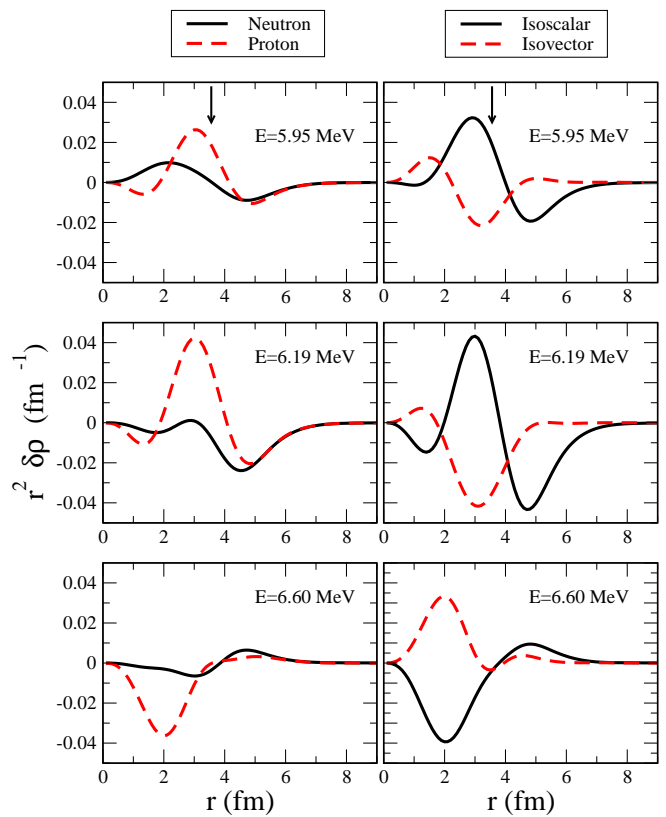} 
\caption{\small (Color online) Neutron and proton transition densities (left) and
the corresponding isoscalar and isovector ones (right) associated to the peaks located at 5.95 (upper panels), 6.19
(middle panels) and 6.60 (lower panels) MeV.}
\label{Trd-1}
\end{flushleft}
\end{figure}

\begin{figure}
\begin{flushleft}
\epsfig{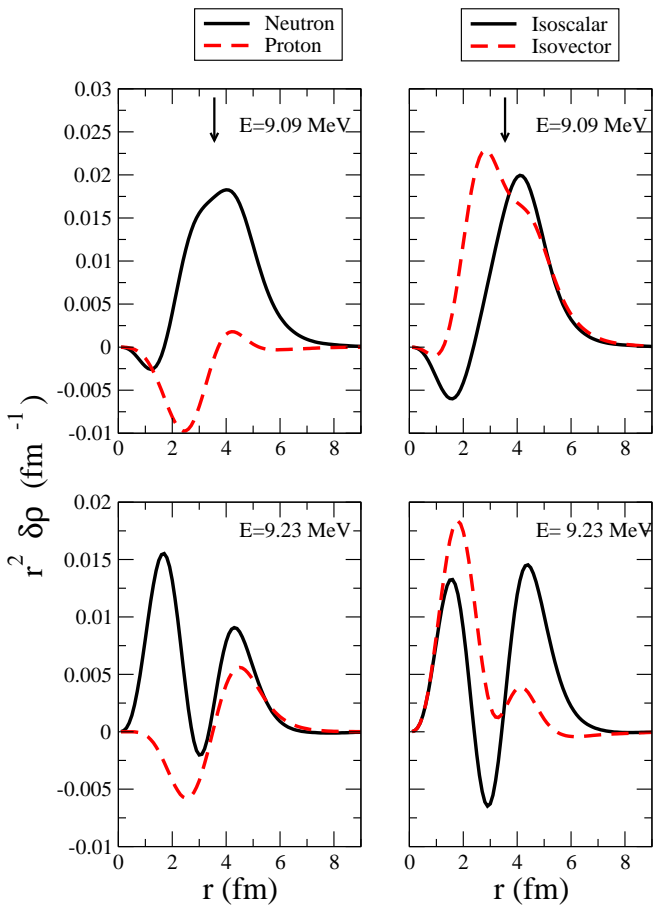}
\caption{\small (Color online) As in Fig. \ref{Trd-1} but for the states located 
 at 9.09 and 9.23 MeV (upper and lower panels, respectively). }
\label{Trd-2}
\end{flushleft}
\end{figure}

The nature of the low-energy peaks can be better analyzed by looking at 
the associated transition densities. In Fig. \ref{Trd-1} we compare
the transition densities corresponding to the peaks of 5.95 (upper panels), 6.19
(middle panels) and 6.60 (lower panels) MeV. For each state, we show separately
the neutron $ \delta \rho_n(r)$
and the proton $\delta \rho_p(r)$ (left), 
the isoscalar
$\delta \rho_n(r) + \delta \rho_p(r)$ and the isovector $\delta \rho_n(r) - \delta \rho_p(r)$ 
(right) transitions densities.
The same quantities are shown in Fig. \ref{Trd-2} for the states
located at 9.09 (upper panels) and 9.23 (lower panels) MeV.

In Table \ref{tab-01} we report for each state the isoscalar $B(E1,T=0)$ and
isovector $B(E1,T=1)$ transition probabilities obtained by integrating the corresponding
sets of Figs \ref{Trd-1} and \ref{Trd-2} multiplied by $r$.
By looking at the transition densities of the first three states (Fig.
\ref{Trd-1}) one does not see any clear signature of an oscillation of the neutron skin
against the core at the surface of the nucleus. On the contrary, especially in the external 
part of the nucleus, the  protons and neutrons oscillate in phase.
For the state located at 5.95 MeV, the isoscalar and 
isovector transition densities strongly oscillate giving almost
vanishing $B(E1,T=0)$ and $B(E1,T=1)$ values 
(see Table \ref{tab-01}). A similar behavior is found for the state of 6.60 MeV.
For the  state lying at 6.19 MeV the cancellations are less important.
Strong cancellations occur for the state located at 9.23 MeV, (lower panel of Fig. \ref{Trd-2}),
resulting in very small $B(E1,T=0)$ and $B(E1,T=1)$ values (Table \ref{tab-01}).
A different situation is found for the most collective state located at 9.09 MeV. We see that the neutron transition density dominates over the proton one that is almost vanishing in the external part of the nucleus while in the interior the two densities oscillate out of phase. This leads to a strong mixing of 
isoscalar and isovector components.

\begin {table} 
 \begin{tabular}{|c|c|c|}

\hline 
$E$ (MeV) & $B(E1,T=0)$   & $B(E1,T=1)$ \\
 \hline  

5.95& 0.002 & 0.013 
\\
6.19& 0.176 & 0.157 
\\
6.60& 0.011 & 0.036 
\\
9.09& 0.099 & 0.186 
\\
9.23& 0.042 & 0.012
\\
\hline 
\end{tabular}
 \caption {\label{tab-01} Isoscalar $B(E1,T=0)$ ($e^2 fm^2$) and
isovector $B(E1,T=1)$ ($e^2 fm^2$) transition probabilities obtained by integrating the corresponding
curves shown in Figs \ref{Trd-1} and \ref{Trd-2} multiplied by $r$.  }
\end{table}

\begin {table} 
  
\begin{tabular}{|ccccc|}
\hline
&& $E=$ 5.95 MeV &&\\
\hline 
$ph$ conf. & $E$ (MeV)   & $A_{ph}$& $b_{ph} (E1)$&$F_{ph}^\lambda$\\
 \hline  
$(2p_{3/2},1d_{5/2})^\pi$  &   17.499  &    0.001  &    0.021  &    1.343
 \\
$(1f_{7/2},1d_{5/2})^\pi$  &   11.732  &    0.062  &   -0.470  &    3.304
 \\
$(2p_{3/2},2s_{1/2})^\pi$  &   12.444  &    0.022  &    0.140  &    1.726
 \\
$(2p_{1/2},2s_{1/2})^\pi$  &   14.133  &    0.004  &    0.043  &   -1.233
 \\
$(2p_{3/2},1d_{3/2})^\pi$  &   12.120  &    0.172  &    0.107  &    0.451
 \\
$(2p_{1/2},1d_{3/2})^\pi$  &   13.809  &    0.007  &   -0.050  &    1.053
 \\
$(1f_{5/2},1d_{3/2})^\pi$  &   13.867  &    0.012  &    0.188  &    2.756
 \\
$(2p_{3/2},2s_{1/2})^\nu$ &   11.773  &    0.006  &    0.060  &    1.737
 \\
$(2p_{1/2},2s_{1/2})^\nu$ &    13.556 &     0.001 &    -0.011 &    -1.248
 \\
$(2p_{3/2},1d_{3/2})^\nu$ &    10.329 &     0.145 &    -0.071 &     0.456
 \\
$(2p_{1/2},1d_{3/2})^\nu$ &    12.112 &     0.001 &     0.014 &     1.084
 \\
\hline 
Partial Sum&&                                       0.433&     -0.031&\\        
 
Total Sum&&                                       0.435&     -0.054&\\
\hline
\end{tabular}
\caption {\label{tab-1} Particle-hole configurations which give
    the major contributions to the dipole low-lying state located at 5.95 MeV. For
    each $ph$ configuration, the energy,
    the contribution to the norm of the state $A_{ph}$, the partial
    contribution to the reduced transition amplitude $b_{ph}$ ($e$
    fm) and the matrix element of the transition operator $F_{ph}^\lambda$
are reported. The superscripts $\pi$, $\nu$ refer to 
    proton and neutron states, respectively.}
\end{table}

 \begin {table} 
 
\begin{tabular}{|ccccc|}
\hline
&& $E=$ 6.19 MeV &&\\
\hline 
$ph$ conf. & $E$ (MeV)   &$A_{ph}$& $b_{ph} (E1)$&$F_{ph}^\lambda$\\
 \hline  
$(1f_{7/2},1d_{5/2})^\pi$ &      11.732   &     0.156   &    -0.727 &      
3.304
 \\
$(2p_{3/2},2s_{1/2})^\pi$ &      12.444   &     0.148   &     0.373 &      
1.726
 \\
$(2p_{1/2},2s_{1/2})^\pi$ &      14.133   &     0.003   &    -0.043 &     
-1.233
 \\
$(2p_{3/2},1d_{3/2})^\pi$ &      12.120   &     0.003   &    -0.015 &      
0.451
 \\
$(2p_{1/2},1d_{3/2})^\pi$ &      13.809   &     0.073   &     0.166 &      
1.053
 \\
$(1f_{5/2},1d_{3/2})^\pi$ &      13.867   &     0.018   &     0.250 &      
2.756
 \\
$(2p_{3/2},1d_{5/2})^\nu$&       15.683  &      0.000  &      0.012&       
1.410
 \\
$(2p_{3/2},2s_{1/2})^\nu$&       11.773  &      0.015  &     -0.079&       
1.737
 \\
$(2p_{3/2},1d_{3/2})^\nu$&       10.329  &      0.040  &      0.038&       
0.456
 \\
$(2p_{1/2},1d_{3/2})^\nu$&       12.112  &      0.001  &     -0.016&       
1.084
 \\
$(1g_{9/2},1f_{7/2})^\nu$&       11.364  &      0.003  &     -0.106&       
4.171
 \\
\hline 
 Partial Sum&&                                     0.461 &    -0.147 &\\
 Total Sum&&                                       0.465 &    -0.163&   \\
\hline
\end{tabular}
 \caption {\label{tab-2} As in Table \ref{tab-1} but for the state located at 6.19 MeV}
\end{table}

Another interesting analysis that can be done for these excitation modes is
related to their collectivity in terms of number and coherence of the 
different $1p1h$
configurations which contribute the total transition probability. We 
present an analysis similar to that done in Ref. \cite{lanza}.
In SRPA as well as in RPA the reduced transition probability for a one-body
operator describing the excitation from the ground state to a  
state $\nu$ can be written as
\begin{equation}
B(E\lambda) = |\sum_{ph} b_{ph}(E\lambda)|^2 =|\sum_{ph} (X_{ph}^\nu -
Y_{ph}^\nu)
 F_{ph}^\lambda|^2
 \label{bel}
\end{equation}
where $F_{ph}^\lambda$ are the multipole transition amplitudes associated 
to a $1p1h$ configuration.
We remark that also in the case of SRPA only $1p1h$ 
amplitudes appear in the expression of the transition probability. A different situation would occur if
a two-body operator is considered. In Refs. \cite{ni1} and \cite{roth}, for  example, the study of
the double giant dipole resonance in $^{40}$Ca and $^{16}$O has been 
carried out by using a two-body operator. In the spirit of a multiphonon picture, the latter is built as a
product of two one-body dipole operators. As shown in Fig. \ref{Ca48N1}, the
low-lying dipole states that we obtain in SRPA have a strong $2p2h$ nature. It  could thus be
interesting to investigate their properties by using a two-body transition
operator. In particular, the use of a transition operator containing both  one-body and two-body
terms is expected to affect the strength distribution and eventually the total strength
associated to this energy region. On the other hand, it is not clear which kind of one-body
multipole operators should be taken into account here to construct the two-body
operator. This investigation is left as a subject for a future work.

In Tables \ref{tab-1}-\ref{tab-5} we report  
the particle-hole configurations which provide the major contributions to the dipole
modes for the five states analyzed in Figs \ref{Trd-1} and \ref{Trd-2}. For each configuration we report the unperturbed energy, the
contribution $A_{ph}$ to the norm of the state,
\begin{equation}\label{aph}
A_{ph} = |X^\nu_{ph}|^2 -|Y^\nu_{ph}|^2,  ~~~\mbox{with}~~~ \sum_{ph}A_{ph} =
N_1, 
\end{equation}
 the partial contribution $b_{ph}$ to the reduced transition amplitude (see Eq.
(\ref{bel})) 
 and the matrix element of the transition operator. 

\begin {table} 
  
\begin{tabular}{|ccccc|}
\hline
&& $E=$ 6.60 MeV &&\\
\hline 
$ph$ conf. & $E$ (MeV)   &$A_{ph}$& $b_{ph} (E1)$&$F_{ph}^\lambda$\\

 \hline  
$(2p_{3/2},1d_{5/2})^\pi$ &    17.499  &     0.001 &     0.031   &   1.343
 \\
$(1f_{7/2},1d_{5/2})^\pi$ &    11.732  &    0.012  &    0.226    &  3.304
 \\
$(1f_{5/2},1d_{5/2})^\pi$ &    19.246  &    0.001  &    0.010    & -0.734
 \\
$(2p_{3/2},2s_{1/2})^\pi$ &    12.444  &    0.190  &    0.421    &  1.726
 \\
$(2p_{1/2},2s_{1/2})^\pi$ &    14.133  &    0.002  &   -0.049    & -1.233
 \\
$(2p_{1/2},1d_{3/2})^\pi$ &    13.809  &    0.335  &   -0.350    &  1.053
 \\
$(1f_{5/2},1d_{3/2})^\pi$ &    13.867  &    0.014  &   -0.193    &  2.756
 \\
$(2p_{3/2},2s_{1/2})^\nu$&     11.773 &     0.005 &    -0.044   &   1.737
 \\
$(2p_{3/2},1d_{3/2})^\nu$&     10.329 &     0.035 &     0.035   &   0.456
 \\
$(2p_{1/2},1d_{3/2})^\nu$&     12.112 &     0.018 &    -0.060   &   1.084
 \\
$(2d_{5/2},1f_{7/2})^\nu$&     12.698 &     0.001 &     0.015   &   1.212
 \\
$(3d_{5/2},1f_{7/2})^\nu$ &    15.625  &    0.000  &    0.010   &   1.055
 \\
$(1g_{9/2},1f_{7/2})^\nu$ &    11.364  &    0.001  &    0.058   &   4.171
 \\
\hline 
  Partial Sum&&                                         0.615 &      0.110 &\\
  Total Sum&&                                           0.622&       0.104 &\\
\hline
\end{tabular}
\caption {\label{tab-3} As in Table \ref{tab-1} but for the state located at 6.60 MeV}
\end{table}

\begin{figure}
\begin{flushleft}
\epsfig{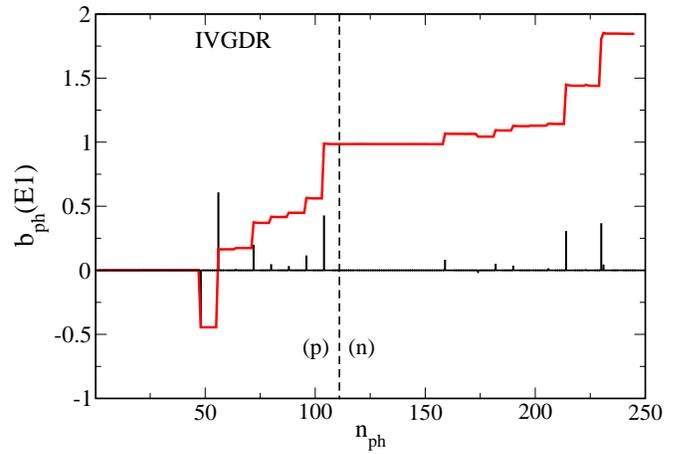} 
\caption{(Color online)
 Partial contributions $b_{ph}$ of the
  reduced transition probability versus the ordering number of the $1p1h$
  configurations for the first peak of the RPA IVDGR located at 17.33 MeV (Fig. \ref{RPA-Ca48}).
The dashed line separates the proton from the neutron
  configurations. 
The configurations are ordered according to their increasing energy.
The bars corresponds to the individual $b_{ph}$ contributions
  while the full red line is the cumulative sum of the
  contributions. \label{Coll-gdr} }

\end{flushleft}
\end{figure}

We first briefly
discuss the case of the RPA IVGDR whose collective features are  well known.
In Fig. \ref{Coll-gdr} the partial contributions $b_{ph}$ 
corresponding to each $1p1h$
configuration for the first  peak  located at 17.33 MeV (Fig. \ref{RPA-Ca48}) are shown. The bars correspond to each 
value of $b_{ph}$ associated to a single configuration while the continuous line
is the cumulative sum of the contributions. The dashed line separates the
proton from the neutron configurations which are ordered according to their
increasing energy. 
We can clearly see  that the contributions of many proton and neutron $1p1h$ configurations 
sum up coherently to provide the total $B(E1)$. 
This coherent behavior of protons and neutrons is due to the minus sign in the
definition of the isovector transition operator, Eq. (\ref{trans-operator}), and is not in contrast with
the isovector character of this excitation where neutrons and protons
oscillate out of phase. 

\begin{figure}
\begin{flushleft}
\epsfig{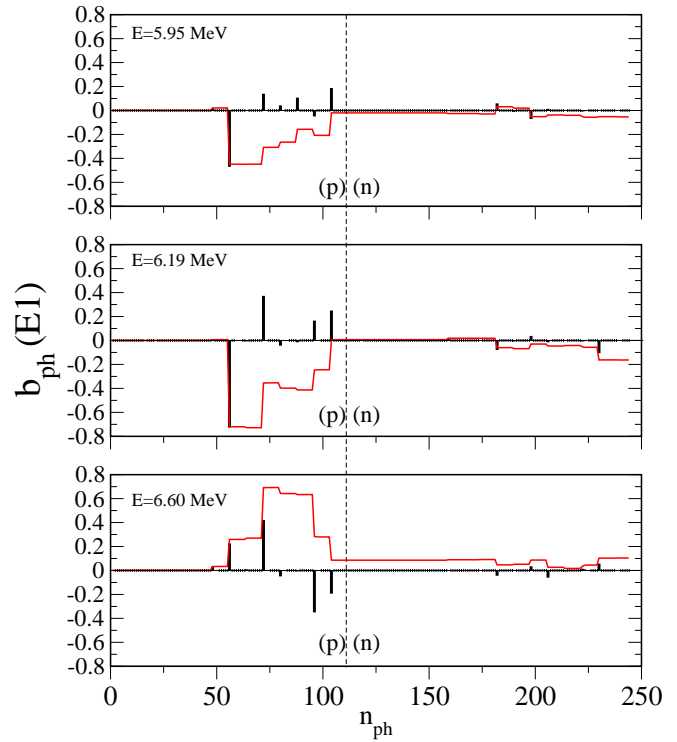} 
\caption{(Color online)
 As in Fig. \ref{Coll-gdr} but for the state at
at 5.95 MeV, 6.19 MeV and 6.60 MeV.   }
\label{Coll-1}
\end{flushleft}
\end{figure}

\begin{figure}
\begin{flushleft}
\epsfig{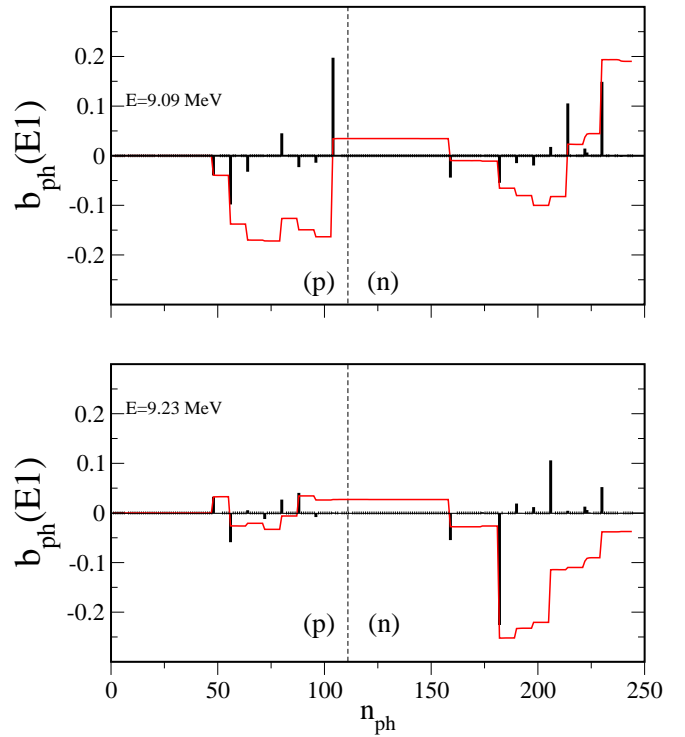} 
\caption{\small (Color online)
 As in Fig. \ref{Coll-1} but for the state at 9.09 and 9.23 MeV (upper and lower
panel, respectively). Please note that a different scale has been used in the ordinate with
respect to Fig. \ref{Coll-1}.}
\label{Coll-2}
\end{flushleft}
\end{figure}

In Figs. \ref{Coll-1} and \ref{Coll-2} the partial contributions $b_{ph}$ 
corresponding to each $1p1h$
configuration for the excitation modes located at
5.95, 6.19 and 6.60 MeV (Fig. \ref{Coll-1}) and at 9.09 and 9.23 MeV (Fig.
9) are plotted.

In the upper panel of Fig. \ref{Coll-1} we see that, for the state lying at 5.95
MeV, several $1p1h$ configurations present non negligible $b_{ph}$ values. This
is true especially  for the proton configurations and is also indicated by the
behavior of the proton
transition density in Fig. \ref{Trd-1}. However, the total amplitude is very
small since strong cancellations occur. The same holds for the state located at
6.19 MeV (middle panel of the same figure), but only for the proton $1p1h$
configurations whose cumulative sum is almost zero. The total transition 
amplitude is
given in this case only by the neutron configurations.  
A different result is found for the third state shown in the lower panel, where
the strong cancellation occurs for the neutron  configurations. 
For the highest state lying at 9.23 MeV (lower panel of Fig. \ref{Coll-2}) we observe
strong cancellations for both neutrons and protons leading to a very small
total transition amplitude.
A more interesting situation is obtained for the most collective state, located
at 9.09 MeV (upper panel of Fig. \ref{Coll-2}). We observe also in this case a
strong cancellation of the proton contributions while a quite coherent behavior
is exhibited by the neutron $1p1h$ configurations. In particular, we observe
strong contributions coming from the  outermost  neutrons (see also Table
\ref{tab-4}). This result, together with the profile of the corresponding 
transition density indicates that this state shows some features usually associated to 
pygmy resonances. 
\begin {table} 
 
\begin{tabular}{|ccccc|}
\hline
&& $E=$ 9.09 MeV &&\\
\hline 
$ph$ conf. & $E$ (MeV)   &$A_{ph}$& $b_{ph} (E1)$&$F_{ph}^\lambda$\\
 \hline  
$(2p_{3/2},1d_{5/2})^\pi$ &     17.499 &       0.002 &      -0.039 &       1.343
 \\
$(1f_{7/2},1d_{5/2})^\pi$ &     11.732 &       0.001 &      -0.098 &       3.304
 \\
$(1f_{5/2},1d_{5/2})^\pi$ &     19.246 &       0.006 &      -0.032 &      -0.734
 \\
$(2p_{1/2},2s_{1/2})^\pi$ &     14.133 &       0.005 &       0.045 &      -1.233
 \\
$(2p_{3/2},1d_{3/2})^\pi$ &     12.120 &       0.007 &      -0.023 &       0.451
 \\
$(2p_{1/2},1d_{3/2})^\pi$ &     13.809 &       0.000 &      -0.014 &       1.053
 \\
$(1f_{5/2},1d_{3/2})^\pi$ &     13.867 &       0.017 &       0.197 &       2.756
 \\
$(2p_{3/2},1d_{5/2})^\nu$&      15.683&        0.006&       -0.044&        1.410
 \\
$(2p_{3/2},2s_{1/2})^\nu$&      11.773&        0.006&       -0.055&        1.737
 \\
$(2p_{1/2},2s_{1/2})^\nu$ &     13.556 &       0.001 &      -0.015 &      -1.248
 \\
$(2p_{3/2},1d_{3/2})^\nu$ &     10.329 &       0.011 &      -0.020 &       0.456
 \\
$(2p_{1/2},1d_{3/2})^\nu$ &     12.112 &       0.001 &       0.018 &       1.084
 \\
$(1f_{5/2},1d_{3/2})^\nu$ &     14.072 &       0.009 &       0.105 &       2.675
 \\
$(2d_{5/2},1f_{7/2})^\nu$ &     12.698 &       0.001 &       0.014 &       1.212
 \\
$(1g_{9/2},1f_{7/2})^\nu$ &     11.364 &       0.008 &       0.149 &       4.171
 \\
\hline 
  Partial Sum&&                                         0.082 &      0.189 &\\
  Total Sum&&                                         0.083 &      0.190   &\\
\hline                     
\end{tabular}
 \caption {\label{tab-4} As in Table \ref{tab-1} but for the state of energy 9.09 MeV}
\end{table}               

\begin {table} 
 
\begin{tabular}{|ccccc|}
\hline
&& $E=$ 9.23 MeV &&\\
\hline 
$ph$ conf. & $E$ (MeV)   &$A_{ph}$& $b_{ph} (E1)$&$F_{ph}^\lambda$\\
 \hline  
$(2p_{3/2},1d_{5/2})^\pi$  &     17.499   &     0.002  &      0.033  &     
1.343
 \\
$(1f_{7/2},1d_{5/2})^\pi$  &     11.732   &     0.001  &     -0.059  &     
3.304
 \\
$(2p_{3/2},2s_{1/2})^\pi$  &     12.444   &     0.000  &     -0.012  &     
1.726
 \\
$(2p_{1/2},2s_{1/2})^\pi$  &     14.133   &     0.001  &      0.027  &    
-1.233
 \\
$(2p_{3/2},1d_{3/2})^\pi$  &     12.120   &     0.023  &      0.040  &     
0.451
 \\
$(2p_{3/2},1d_{5/2})^\nu$ &      15.683  &      0.008 &      -0.055 &      
1.410
 \\
$(2p_{3/2},2s_{1/2})^\nu$ &      11.773  &      0.103 &      -0.226 &      
1.737
 \\
$(2p_{1/2},2s_{1/2})^\nu$ &      13.556  &      0.001 &       0.019 &     
-1.248
 \\
$(2p_{3/2},1d_{3/2})^\nu$ &      10.329  &      0.004 &       0.012 &      
0.456
 \\
$(2p_{1/2},1d_{3/2})^\nu$  &     12.112   &     0.056  &      0.106  &     
1.084
 \\
$(2d_{5/2},1f_{7/2})^\nu$  &     12.698   &     0.001  &      0.013  &     
1.212
 \\
$(1g_{9/2},1f_{7/2})^\nu$  &     11.364   &     0.001  &      0.052  &     
4.171
 \\
\hline 
  Partial Sum&&                                         0.201&     -0.050   &\\
  Total Sum&&                                        0.202  &    -0.037      &\\
\hline                      
\end{tabular}
 \caption {\label{tab-5} As in Table \ref{tab-1} but for the state lying at 9.23 MeV}
\end{table} 

\begin {table}

  \begin{center}
\begin{tabular}{|c|c|c|c|}
\hline
&&$^{48}$Ca&$^{40}$Ca\\
\hline
$\sum B(E1)$&SRPA&230 &23\\
(10$^{-3}$ e$^{2}$ fm$^{2}$)& Exp& 68.7 $\pm$ 7.5     & 5.1 $\pm$ 0.8     \\
\hline

$\sum_i E_i B_i(E1)$&SRPA&1964&211\\
10$^{-3}$  e$^{2}$ fm$^{2}$ MeV      &Exp &570 $\pm$ 62      &35 $\pm$ 5      \\ 
\hline
$E_{centroid}$ &SRPA &8.54 &9.17\\
MeV&Exp&8.40&6.80 \\
\hline	

\end{tabular}
\caption{\label{Tab-Th-Exp}
Total $B(E1)$ and EWSRs integrated up to 10 MeV and corresponding centroid energies obtained in SRPA compared with the experimental values \cite{tre-1} for the $^{40,48}$Ca isotopes.} 
\end{center}
\end{table}

\begin{figure}
\begin{flushleft}
\epsfig{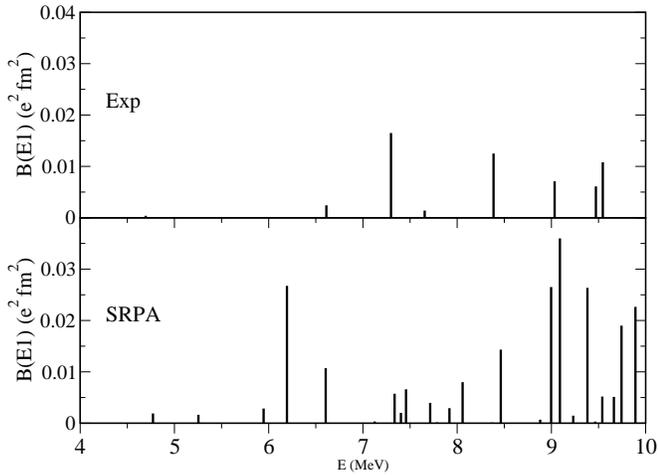} 
\caption{\small 
Comparison of the experimental B(E1) strength 
distribution \cite{tre-1} (upper panel) with the SRPA calculations (lower panel) for $^{48}$Ca.}
\label{Th-Exp}
\end{flushleft}
\end{figure}

In Table \ref{Tab-Th-Exp} we compare the total $B(E1)$ and EWSRs integrated up to 10 MeV and the corresponding centroid energies   with the experimental values \cite{tre-1} for the nuclei $^{40}$Ca and $^{48}$Ca. We see that the SRPA total $B(E1)$ is much larger, almost by a factor 4, than the experimental value. This is due to the larger number of states obtained in SRPA with respect to the experimental spectrum and, at the same time, to their higher strength (Fig. \ref{Th-Exp}). The same kind of discrepancies is found for the EWSRs while the SRPA centroid energy is very close to the experimental value. Regarding these large differences some comments are in order. First we recall that, because of the non-local terms of the Skyrme interaction, the double commutator sum rule is enhanced with respect to the  classical  sum rule by a factor 1.35 for SGII. However, this is not enough to explain the 
strong deviation with respect to the experimental values. It could be also interesting to analyze whether or not this kind of discrepancy may depend on the choice of the Skyrme interaction. As a check, we have performed  SRPA calculations by using 
the parametrization 
SLy4 and the same kind of deviations has been found. In a recent work \cite{colo} it has been suggested that the low-lying dipole strength distribution could be be eventually related to the slope of the symmetry energy. 
This kind of analysis is however beyond our present scopes and it will be performed in future investigations. Finally, we recall that theoretical $B(E1)$ values much larger than the corresponding experimental results have also been found within the ETFFS model for the nucleus $^{44}$Ca; it has been shown that this discrepancy is related to the use of some approximations in the treatment of pairing and continuum coupling \cite{tertychny}. 
We also mention that in previous experimental measurements a much larger strength had been found in the energy region from 5 to 10 MeV \cite{Ottini}. 
Finally, as mentioned above,  since the low-lying dipole states obtained
in SRPA have a strong $2p2h$ nature, it could more appropriate the use of a
transition operator containing both one-body and two-body terms. Of course, the
use of such a more general operator would affect the total strength associated
to this energy region. This investigation is however left as a subject for a
future work.

\section{Mixing with the spurious state}

Some comments about the spurious state and its possible mixing with the physical
 dipole modes are in order. The Thouless theorem on the EWSRs \cite{thou} is
very important in the framework of RPA and it holds also in SRPA \cite{yannou}.
It guarantees that spurious excitations corresponding to some symmetries 
separate out and are orthogonal to the physical states. This separation is
obtained only in completely self-consistent calculations, that is, when  the
same interaction is used at both the HF and the RPA (or SRPA) level. In the case
of dipole excitations, the center-of-mass motion should appear at zero energy 
and the EWSRs should be satisfied. Since the Coulomb and spin-orbit terms are
not taken into account in the residual interaction, our  calculations are  not
fully self-consistent  and violations of the  EWSRs are found (not larger than
$2-3\%$). A currently adopted procedure to estimate the mixing with
spurious components consists in using the isoscalar one-body operator corrected for
the center-of-mass motion. This has been done, for example, in the study of giant resonances in Ref. \cite{gamba2010-1}.
However, this check which is generally employed in RPA calculations is 
suitable to analyze excitations which are 
mainly composed by $1p1h$ states. In our case, as shown in Fig.~\ref{Ca48N1},
$1p1h$ and $2p2h$ components are strongly mixed and we thus
 choose a more appropriate procedure to estimate the mixing with spurious components.

\begin{figure}
\begin{flushleft}
\epsfig{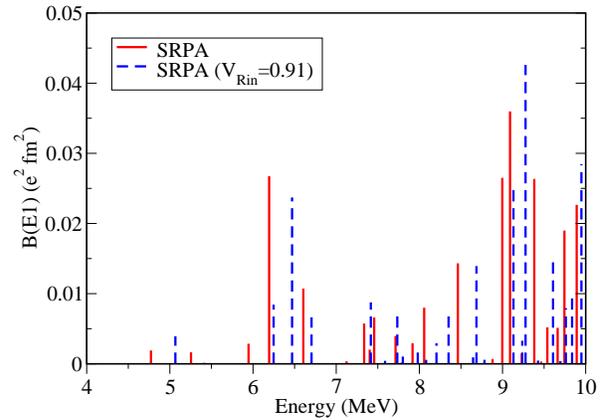}
\caption{\small (Color online) Dipole strength distribution for the $^{48}$Ca
isotope obtained in SRPA without (full red lines) and with a renormalizing factor $V_{Rin}=0.91$ in the residual interaction (dashed blue lines). Both calculations are done with an energy cutoff  on the $2p2h$ configurations of 60 MeV.}
\label{spurious}
\end{flushleft}
\end{figure}

The residual interaction has been multiplied by a
renormalizing factor to shift to zero the energy of the spurious mode.
In our RPA calculations, the
spurious state lies at about 3.5 MeV  exhausting more than 95\% of the isoscalar
EWSRs; by using a renormalizing factor of 1.09 its energy goes down to 0.2 MeV
while the rest of the isoscalar and isovector distributions remains practically
unaffected.  In SRPA, as a consequence of the coupling with the $2p2h$
configurations, the spurious state is pushed down to lower energies with respect
to RPA. Depending on the energy 
cutoff on the $2p2h$ configurations, its energy can be negative or imaginary in
some cases. 
We stress that the use of the Hartree-Fock ground state 
that minimizes the energy 
guarantees that the RPA matrix is positive definite. This does not hold in
SRPA. 
For the SRPA calculations with $ECUT=$ 60 MeV 
we have verified that, 
by using a renormalizing factor of 0.91, the spurious mode is located at about
0.12 MeV. The total $B(E1)$ (EWSRs)  integrated up to 10 MeV does not change
much when the renormalizing factor is introduced: from  0.230 $e^2$ fm$^2$ 
(1.964 $e^2$ fm$^2$ MeV) to 0.221 $e^2$ fm$^2 $ (1.905 $e^2$ fm$^2$ MeV) . In 
Fig. \ref{spurious} we show the
dipole strength distribution obtained for the $^{48}$Ca 
isotope without (full red lines) and with (dashed blue lines) the inclusion of
the renormalizing factor in 
the residual interaction. Both calculations are done with an energy cutoff  on
the $2p2h$ configurations of 60 MeV. We see that the strength distribution is
not strongly affected, the main difference being a shift of few hundreds KeV of
some peaks and a change of their corresponding transition probability. This
analysis seems to indicate that the presence of possible
spurious components does not affect much our results.   

\section{Conclusions}

In this work, we have analyzed the low-energy dipole spectrum (from 5 to 10 MeV)
for the stable nuclei $^{40}$Ca and $^{48}$Ca in the framework of the
Skyrme-SRPA model. The Skyrme interaction SGII is used. Almost no strength is
found in this energy region for the nucleus $^{40}$Ca whereas a non negligible
strength is obtained for the neutron-rich nucleus $^{48}$Ca. The distribution
and the fragmentation of the peaks is in reasonable agreement with the
corresponding experimental measurements. This kind of results cannot be provided
by the standard RPA model: SGII-RPA calculations do not lead to any strength in
the energy region from 5 to 10 MeV. However, we have found a $B(E1)$ value
integrated up to 10 MeV which is quite larger than the corresponding
experimental result.  

The inclusion of $2p2h$ configurations and their coupling with the $1p1h$ ones
and among themselves within SRPA has a two-fold effect on the low-lying dipole
strength:
(i) states that already exist in RPA are  shifted 
to lower energies. These states maintain a quite strong $1p1h$ nature; (ii) 
several other states of almost pure $2p2h$ character appear. They are excited by
the one-body dipole operator through their $1p1h$ components. In the nucleus 
 $^{48}$Ca one of these states shows the largest transition strength and
displays some features generally associated with a dipole pygmy resonance.

A detailed analysis of the main excitations that compose the strength
distributions is done: the content of $1p1h$ and $2p2h$ configurations  is
studied for some peaks. The transition densities are shown and the collectivity
of the peaks is investigated in terms of the number and the coherence of the
single configurations that mainly contribute. 
As far as the transition densities are concerned, one may conclude that in
general they do not display the typical profile well known for pygmy resonances,
except for the state located at 9.09 MeV. 
We have also observed that about 10-12 configurations contribute to each peak.
In this sense one can say that there is some collectivity, however, strong
cancellations occur in most cases and the single configurations do not sum up in
a coherent way. 

Our main conclusion is that, even if a low-lying dipole response is found
experimentally and predicted theoretically in the isotope $^{48}$Ca, one cannot
really describe these excitations as pygmy resonances except for the most
collective peak located at 9.09 MeV. This suggests that this nucleus is still
too light to present clear signatures of an oscillation of the neutron skin
against the internal core and that individual degrees of freedom are still
dominant in the description of the dipole low-energy spectrum as for lighter
nuclei.

\end{document}